\documentclass[prl,twocolumn,aps,showpacs]{revtex4}
\usepackage{epsf,epsfig,graphicx}
\usepackage{amsmath}
\usepackage{subfigure}
\begin{document}
\title{Disorder-Induced Shift of Condensation Temperature for Dilute Trapped Bose Gases}
\author{Matthias Timmer}
\email{sp0088@theo-phys.uni-essen.de}
\affiliation{Universit{\"a}t Duisburg-Essen, Campus Essen, Fachbereich Physik, 
Lotharstrasse 1, 47048 Duisburg, Germany}
\author{Axel Pelster}
\email{axel.pelster@uni-due.de}
\affiliation{Universit{\"a}t Duisburg-Essen, Campus Essen, Fachbereich Physik, 
Lotharstrasse 1, 47048 Duisburg, Germany}
\author{Robert Graham}
\email{robert.graham@uni-due.de}
\affiliation{Universit{\"a}t Duisburg-Essen, Campus Essen, Fachbereich Physik, 
Lotharstrasse 1, 47048 Duisburg, Germany}
\date{\today}
\begin{abstract}
We determine the leading shift of the Bose-Einstein condensation
temperature for an ultracold dilute atomic gas in a harmonic trap due to weak disorder 
by treating both a Gaussian and a Lorentzian spatial correlation for the quenched
disorder potential. Increasing the correlation length from values
much smaller than the geometric mean of the trap scale and the mean
particle distance to much larger values
leads first to an increase of the positive shift to a maximum at this critical length
scale and then to a decrease.
\end{abstract}
\pacs{03.75.Hh,64.60.Cn}
\maketitle
The experimental realization of Bose-Einstein condensates (BECs) in magnetically
trapped dilute atomic vapors \cite{Cornell,Ketterle1} has renewed the 
theoretical interest in the phase structure of ultracold Bose 
gases. In particular, many research groups have investigated  
how the transition temperature $T_c$
changes due to
a weak local repulsive two-particle interaction
$V^{ ({\rm int}) }({\bf x})=4 \pi a / M\,\delta ({\bf x})$ 
in different trap configurations. Here $a$ denotes the s-wave scattering length,
$M$ the atomic mass, and we employ units with $k_B= \hbar=1$.
Theoretically most demanding is the homogeneous case \cite{Andersen}
where the Bose gas of particle density $n$ 
resides in an infinitely wide trap with a flat potential. There the leading
shift $\Delta T_c / T_c^{(0)} = c_1 a n^{1/3}$
with respect to the interaction-free critical temperature
$T_c^{(0)}= 2 \pi \left[ n / \zeta (3/2) \right]^{2/3}/ M$
is completely due to long-wavelength, critical fluctuations that
have to be described nonperturbatively. It can be shown that this shift
follows from the classical limit of many-body
theory, where only the zero Matsubara modes of the Bose fields are
included \cite{Baym1}. A self-consistent variational resummation \cite{Kleinertbec} 
of the resulting infrared divergent series for $c_1$ 
yields the value $1.27 \pm 0.11$ \cite{Boris2} which
agrees with the Monte Carlo calculations
$1.32\pm0.02$ \cite{Svistunov,Arnold1}.
In contrast to that, harmonic trapping potentials remove the 
critical long-wavelength fluctuations and reduce the fraction of atoms taking part in
nonperturbative physics at the transition point. As a result, 
the leading shift $\Delta T_c / T_c^{(0)} = d_1 \,a / \lambda_c^{(0)}$
in the critical temperature
with respect to its noninteracting value
$T_c^{(0)}=[N \omega_{\rm g} / \zeta(3)]^{1/3}$
is negative and
can be calculated by simple perturbative methods, for
instance a Hartree-Fock mean-field approximation \cite{Giorgini}.
Here $\lambda_c^{(0)}=\sqrt{2 \pi  / [ T_c^{(0)} M] }$ denotes the
thermal wavelength and $N$ the boson number, 
whereas $\omega_{\rm g} = (\omega_x \omega_y \omega_z )^{1/3}$ is the geometric mean of the
respective trap frequencies.  The analytically known constant $d_1$ 
has the numerical value $d_1 =  - 3.426$ which
could be confirmed in a recent experiment within the error bars \cite{Aspect1}.
Note that a theoretical investigation of the crossover
between both physically distinct cases of the homogeneous and the trapped Bose gas
has been elaborated in Ref.~\cite{Zobay4}
with the help of general power-law potentials.
Furthermore,
corrections to the leading interaction-induced shift in the critical temperature have
been investigated in the Refs.~\cite{Arnold3,Arnold4,Davis,Kao}.\\
Apart from the contact interaction, another interesting physical parameter for
ultracold atomic vapors is frozen disorder. It can be created artificially by
laser speckles \cite{Inguscio1,Aspect2,Inguscio2,Schulte},
incommensurate lattices~\cite{Lewenstein}, or
different localized atomic species \cite{Castin,Sengstock}. 
However, random potentials can also arise naturally via the spatial fluctuations 
of the electric currents generating the magnetic wire traps~\cite{Demler,Aspect3}.
Thus, one can raise the fundamental
question how the Bose-Einstein condensation temperature is affected
by weak disorder. To this end we consider the quite commonly employed model 
of a Gaussian distributed disorder potential $V({\bf x})$ with the ensemble averages
$\overline{V({\bf x})}= 0$ and 
$\overline{V({\bf x}) V({\bf x'})} = R ({\bf x}-{\bf x'})$. In particular, we
are interested in correlation functions $R ({\bf x}-{\bf x'})$ 
which decay with a characteristic correlation length $\xi$. 
A typical example for such a correlation function is a 
Lorentzian distribution, which occurs in atom chips \cite{Schmiedmayer}, more complicated
correlation functions can be designed by using laser speckles \cite{Speckles}.
Note that the latter experiments allow to tune the corresponding correlation length $\xi$
in a wide parameter range by changing, for instance, the distance between the
scattering surface and the magneto-optical trap.
So far, the disorder-induced shift of the critical temperature 
has only been analyzed for a homogeneous Bose gas
in the limit $\xi \to 0$ of a short-range correlation $R ({\bf x}) = R \,\delta ({\bf x})$.
A perturbative calculation of Lopatin and Vinokur \cite{Lopatin}, motivated by
studying the superfluid motion of helium in porous media such as Vycor
\cite{Reppy1,Reppy2,Huang}, yields
\begin{eqnarray}
\label{Hom}
\frac{\Delta T_c}{T_c^{(0)}} = - \frac{M^2 R}{3 \pi [\zeta(3/2)]^{2/3} 
n^{1/3}} \, .
\end{eqnarray}
The negative sign for the homogeneous Bose gas
is compatible with the fact that
disorder leads to an effective attractive interaction
when the replica method is applied \cite{Graham}. \\
The purpose of this note is to determine the leading disorder-induced
shift in the critical temperature for a weakly interacting Bose gas 
in an external harmonic trap $U ( {\bf x} )$ with an additional weak disorder
potential which is characterized
by an arbitrary correlation function  $R ({\bf x}-{\bf x'})$.
According to our introduction
we expect that disorder leads to an opposite sign of the shift 
of the critical temperature than repulsive
interaction. Since disorder appears within the replica method 
as an effective attractive interaction \cite{Graham}, the particle density in the center
of a harmonic trap is increased. 
Thus, there should be a positive disorder-induced shift for such a harmonic
confinement. Moreover, we should be able to determine it reliably
by a perturbative calculation. In first-order perturbation theory,
the effects of both interaction and disorder are additive.
Therefore, we start with the
imaginary-time action of a non-interacting Bose gas 
in both a trap and a random potential in standard notation 
\begin{eqnarray}
{\cal A}[\psi^*,\psi] &=& \int_0^{\beta} d \tau \int d^3 x \,
\psi^* ({\bf x},\tau) \Big[ \,\frac{\partial}{\partial \tau}
- \frac{1}{2 M} {\bf \Delta}  \nonumber \\
&&  + \,U ( {\bf x} ) + V ( {\bf x} ) 
- \mu \Big] \psi ({\bf x},\tau) \, ,
\end{eqnarray}
and  treat the disorder potential $V ( {\bf x} )$ perturbatively.
We aim at determining the particle number $N = - \partial \Omega/ \partial \mu$
from the grand-canonical free energy $\Omega = - \overline{\ln {\cal Z}} / \beta$,
where the partition function is determined by the functional integral
\begin{eqnarray}
{\cal Z}  = \oint {\cal D} \psi^*  \oint {\cal D} \psi \, 
e^{- {\cal A}[\psi^*,\psi] } \, .
\end{eqnarray}
For an ideal Bose gas a semiclassical treatment of the trap potential leads to
\begin{eqnarray}
\label{FF0}
\Omega^{(0)} = \frac{-1}{\beta \lambda^3}\, \int d^3 x \, 
\zeta_{5/2} \left( e^{\beta [\mu- U({\bf x})]} \right) \, ,
\end{eqnarray}
where $\zeta_{a} (z) = \sum_{n=1}^{\infty} z^n/n^a $ is the polylogarithmic function.
Correspondingly, the first-order correction reads
\begin{eqnarray}
&&\Omega^{(1)}
=
-\frac{1}{2 \beta}
\int_{0}^{\beta} d \tau_{1}
\int_{0}^{\beta} d \tau_{2}
\int d^3 x_{1}
\int d^3 x_{2} \nonumber \\ && \hspace*{-0.5cm} \times
R ( {\bf x}_{1} - {\bf x}_{2} )
G^{(0)} ( {\bf x}_{1} , \tau_{1} ; {\bf x}_{2} , \tau_{2} )
G^{(0)} ( {\bf x}_{2} , \tau_{2} ; {\bf x}_{1} , \tau_{1} )\, ,
\label{Omega1}
\end{eqnarray}
where the bosonic Green function has the semiclassical 
Fourier-Matsubara decomposition
\begin{eqnarray}
&& G^{(0)} ( {\bf x}_1 , \tau_1 ; {\bf x}_2 , \tau_2 ) = 
\frac{1}{\beta} \sum_{m=-\infty}^{\infty}
\int \frac{d^3 k}{(2 \pi)^3} \nonumber \\
&& \hspace*{-0.5cm} \times e^{- i \omega_m ( \tau_1 - \tau_2 )
+ i {\bf k} ( {\bf x}_1 - {\bf x}_2) }
\,G^{(0)} \left( \omega_m , {\bf k}; \frac{{\bf x}_1 + {\bf x}_2}{2} \right) 
\end{eqnarray}
with $G^{(0)} ( \omega_m , {\bf k}; {\bf x}) =  \left[ - i \omega_m + {\bf k}^2 / 2 M
+ U ( {\bf x} ) - \mu\right]^{-1}$, ${\bf x}=({\bf x}_1 + {\bf x}_2)/2$, and $\omega_m 
= 2 \pi m / \beta$.
Evaluating in (\ref{Omega1}) the Matsubara series with the Poisson summation
formula \cite{Path} and
the Fourier integrals with the formula of Plemelj \cite{Heitler}, we obtain
\begin{eqnarray}
\label{Omegaresult}
&& \hspace*{-6mm}\Omega^{(1)}
= 
\left(\frac{M}{2 \pi}\right)^{3}
\sum_{n= 1}^{\infty}
\int d^3 x \,
e^{ \beta n [\mu- U({\bf x} )]}
\int d^3 x' \ R ( {\bf x'})
\lim_{\epsilon \searrow 0} \nonumber \\
&& \hspace*{-6mm}\mbox{Im}
\int_{0}^{\infty} d \tau \
\frac{e^{-\epsilon \tau}}{ [\tau (\tau + i \beta n)]^{3/2} }
\exp \left\{- \frac{M \beta n{\bf x'}^{2}}{2 \tau (\tau + i \beta n)} \right\}\,.
\end{eqnarray}
This disorder correction to the grand-canonical free energy is now specialized for
a harmonic trap $U({\bf x})$ and for
typical correlation functions $R({\bf x}) = \int d^3 k / (2 \pi)^3 \,e^{i {\bf k} {\bf x}}
\,R({\bf k})$.
For instance, in case of
a Gaussian $R_{\rm G} ( {\bf k} ) = R \,e^{- {\bf k}^2 \xi^2/2}$
and a Lorentzian $R_{\rm L} ( {\bf k} ) = R /(1 + {\bf k}^2 \xi^2)$,
we find
\begin{eqnarray}
\label{Omega1Gauss}
\Omega^{(1)}_{\rm G}
= - \frac{\beta R}{ 2 (2 \pi)^{3/2} (\beta \omega_{\rm g})^{3}\xi}
\sum_{n=1}^{\infty}
\frac{e^{n \beta \mu}}{n^{2} \left(\xi^{2}+
\frac{\beta n}{4 M}\right)} 
\end{eqnarray}
and, correspondingly,
\begin{eqnarray}
\Omega^{(1)}_{\bf L} & =&  -
\frac{\pi \sqrt{M \beta}R}{2 (2 \pi)^{3/2}\xi^{2}(\beta \omega_{\rm g})^{3}}
\sum_{n= 1}^{\infty}
\frac{e^{\beta n \mu}}{n^{5/2}} \nonumber \\
&& \times \exp \left[\frac{\beta n}{8 M \xi^{2}}
\right]\left[1 -\Phi \left(\sqrt{\frac{\beta n}{8 M \xi^{2}}}
\,\right) \right]\, ,
\label{Omega1Lorenz}
\end{eqnarray}
where we have used for the latter the probability integral 
$\Phi(z)=2/ \sqrt{\pi} \int_0^z d t \,e^{-t^2}$.\\
As follows from the theory of critical phenomena \cite{Zinn-Justin,Verena},
the phase transition to a Bose-Einstein condensate occurs when the 
spatio-temporal integral over the correlation function
\begin{eqnarray}
&&\hspace*{-1.2cm} G( {\bf x}_1 , \tau_1 ; {\bf x}_2 , \tau_2 ) \nonumber \\[3mm]
&& \hspace*{-0.5cm} = \overline{\frac{1}{\cal Z} 
\oint {\cal D} \psi^*  \oint {\cal D} \psi  \,
\psi ( {\bf x}_1 , \tau_1 ) \psi^* ({\bf x}_2 , \tau_2 )
\,e^{- {\cal A}[\psi^*,\psi] }}
\end{eqnarray}
diverges, i.e. $G ( \omega_m , {\bf k} ; {\bf x} )$
becomes infinite for $\omega_m=0$, ${\bf k}={\bf 0}$, and some ${\bf x}$.
This criticality criterion is equivalent to the vanishing of its inverse 
\begin{eqnarray}
G^{-1}( 0 , {\bf 0} ; {\bf x} ) =
G^{(0)\,-1} ( 0 , {\bf 0} ; {\bf x} ) -
\Sigma  ( 0 , {\bf 0} ; {\bf x} )\, ,
\end{eqnarray}
where the  first-order perturbative contribution to the
self-energy is given by
\begin{eqnarray}
\label{Sigma1}
\hspace*{-2mm}\Sigma^{(1)} ( {\bf x}_{1}  , \tau_{1} ; {\bf x}_{2}   , \tau_{2}  )
= R ( {\bf x}_{1} - {\bf x}_{2}   )
G^{(0)} ( {\bf x}_{1}  , \tau_{1} ; {\bf x}_{2}  , \tau_{2} ) \ .
\end{eqnarray}
Therefore, we  define the critical chemical
potential $\mu_c$ from the condition
\begin{eqnarray}
\label{COND}
\begin{array}{@{}c} {\rm Min} \\[-1mm] {\bf x} \end{array}
G^{-1}( 0, {\bf 0} ; {\bf x} ) = 0 \, .
\end{eqnarray}
Up to first order in the disorder we obtain a result which does not depend on the trap
frequencies:
\begin{eqnarray}
\label{MUC}
\mu_c^{(1)} = - 2 M \int d^3 x \, \frac{R({\bf x})}{| {\bf x} |} \, .
\end{eqnarray}
We conclude from (\ref{MUC}) that Bose-Einstein condensation occurs with disorder 
earlier than in the clean case.
Note that the critical chemical potential for the onset
of Bose-Einstein condensation follows from condition (\ref{COND})  also
in other physical situations, for instance, 
when anisotropic
traps are used for rotating Bose gases \cite{Kling}
or when dipolar gases are considered where the
two-particle interaction contains both a short-range, isotropic contact potential
and a long-range, anisotropic dipole-dipole interaction \cite{Dipol1,Dipol2}.\\
\begin{figure}[t]
\begin{center}
\epsfig{file=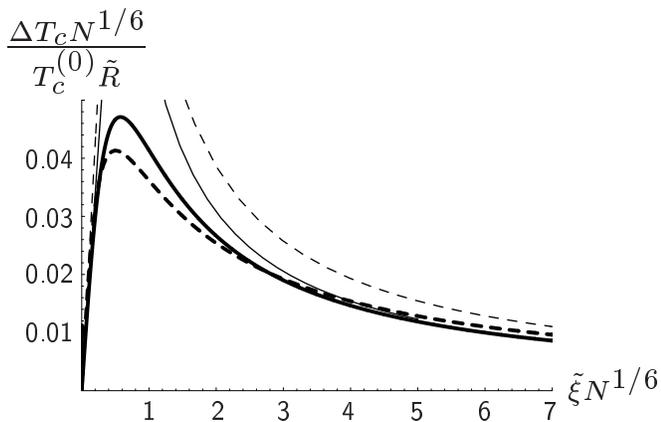,width=\columnwidth} 
\end{center}
\caption{\label{fig1}Disorder-induced shifts (\ref{TcGauss}) and (\ref{TcLorentz})
of the critical temperature $\Delta T_c$ with respect to the
interaction-free critical temperature $T_c^{(0)}$ for the Gaussian (solid) and the 
Lorentzian (dashed) correlation function together with their asymptotics (\ref{ASYM1}) 
and (\ref{ASYM2})
as a function of the dimensionless correlation length.}
\end{figure}
\hspace*{-1.5mm}Now we combine our previous results 
and determine the disorder-induced shift in the 
critical temperature. To this end we evaluate $N = N(\mu)$ in the limit $\mu \uparrow \mu_c$
by taking into account the Robinson formula \cite{Robinson} for
the polylogarithmic function in (\ref{FF0}). For the Gaussian correlation we
find
\begin{eqnarray}
\frac{\Delta T_{c,{\rm G}}}{T_{c}^{(0)}} &=&
\frac{8 \tilde{R} \tilde{\xi}}{3 (2 \pi)^{3/2} \zeta (3)} \nonumber \\
&& \times\sum_{k=1}^\infty
\frac{1}{k^2 \left[ k+ 4 N^{1/3} \tilde{\xi}^2 / [\zeta(3)]^{1/3} \right]} \, ,
\label{TcGauss}
\end{eqnarray}
whereas we obtain for the Lorentzian correlation
\begin{eqnarray}
&&\hspace*{-10mm}\frac{\Delta T_{c,{\rm L}}}{T_{c}^{(0)}}
=
\frac{\tilde{R}}{6 \pi [\zeta(3)]^{2/3}\tilde{\xi}N^{1/3}}
\sum_{k= 1}^{\infty}
\left\{\frac{1}{k^{2}}-
\frac{[\zeta(3)]^{1/6} \sqrt{\pi}}{\sqrt{8} \tilde{\xi}N^{1/6} k^{3/2}}
\right.\nonumber \\
&&\hspace*{-5mm}\left.\times \exp\left[\frac{[\zeta(3)]^{1/3} k}{8 \tilde{\xi}^{2}N^{1/3}}
\right]\left[1 -\Phi\left(
\sqrt{\frac{[\zeta(3)]^{1/3} k}{8 \tilde{\xi}^{2}N^{1/3}}}\,\,
\right)\right]\right\}.
\label{TcLorentz}
\end{eqnarray}
Here we have introduced dimensionless units for the
cor\-relation length $\tilde{\xi}= \xi / l_{\rm os}$
as well as for  the disorder strength $\tilde{R}= M^2 l_{\rm os} R$, 
where $l_{\rm os}= 1/ \sqrt{M \omega_{\rm g}}$ denotes the oscillator length.
Note that both disorder-induced 
shifts (\ref{TcGauss}) and (\ref{TcLorentz}) have similar asymptotics in the limit 
$\xi \to 0$
\begin{eqnarray}
\label{ASYM1}
\frac{\Delta T_{c,{\rm G}}}{T_{c}^{(0)}} \approx  \frac{8 \tilde{R} 
\tilde{\xi}}{3 (2 \pi)^{3/2}} \, , 
\hspace*{0.5cm}
\frac{\Delta T_{c,{\rm L}}}{T_{c}^{(0)}} \approx \frac{2 \tilde{R}\tilde{\xi}}{3 \pi} 
\end{eqnarray}
and in the limit $\xi \to \infty$
\begin{eqnarray}
\frac{\Delta T_{c,{\rm G}}}{T_{c}^{(0)}} 
&\approx& \frac{2 \zeta(2)\tilde{R}}{3 (2 \pi)^{3/2} [\zeta(3)]^{2/3}\tilde{\xi}N^{1/3}}
\, ,  \nonumber \\
\frac{\Delta T_{c,{\rm L}}}{T_{c}^{(0)}} &\approx& \frac{\zeta(2) 
\tilde{R}}{6 \pi [\zeta(3)]^{2/3} \tilde{\xi}N^{1/3}}
\, .
\label{ASYM2}
\end{eqnarray}
\begin{figure}[t]
\begin{center}
\epsfig{file=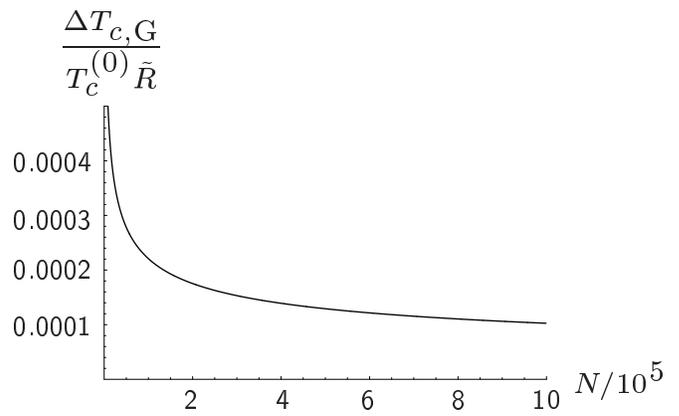,width=\columnwidth} 
\end{center}
\caption{\label{fig2}Normalized shift of the critical temperature 
$\Delta T_{c,{\rm G}} / (T_c^{(0)} \tilde{R})$
of a dilute $^{85}$Rb gas in an anisotropic harmonic trap with geometric mean frequency 
$\omega_{\rm g}=
2 \pi \cdot 40$ Hz and a Gaussian disorder correlation with correlation length 
$\xi = \mbox{10}~\mu$m versus particle number $N$.}
\end{figure}
\hspace*{-1mm}Figure~\ref{fig1} 
shows how the shift results (\ref{TcGauss}) and (\ref{TcLorentz}) 
vary with the dimensionless correlation length $\tilde{\xi}$. Whereas for small 
$\tilde{\xi}$ the
oscillator length $l_{\rm os}$ dominates, the average particle distance
$l_{\rm p}= l_{\rm os}/ N^{1/3}$ becomes important for large $\tilde{\xi}$. Both length scales
meet at $\sqrt{l_{\rm os} l_{\rm p}} = l_{\rm os} / N^{1/6}$, i.e. $\tilde{\xi}N^{1/6}=1$, 
which is roughly where the maximal shift occurs. This result will certainly help to design  
an experiment which aims at detecting a disorder-induced
shift of the critical temperature.\\
Furthermore, it is clear from 
(\ref{TcGauss}), (\ref{TcLorentz}), and Figure~\ref{fig1} that
the shifts for Gaussian or Lorentzian spatial correlations are qualitatively and even
semi-quantitatively the same. This finding justifies to apply our results also to other
correlation functions, which decay with a characteristic correlation length $\xi$
as, for instance, the laser speckle correlations in the 
Florence  experiment \cite{Inguscio1,Inguscio2}. There 
about $N=3\cdot 10^5$ atoms of 
the $^{85}$Rb 
isotope reside in an anisotropic harmonic trap with the geometric mean frequency 
$\omega_{\rm g}=2 \pi \cdot 40$ Hz, so that its oscillator length is 
$l_{\rm os}=\mbox{1.72}~\mu$m.
The disorder potential is characterized by a length scale  
$\xi = \mbox{10}~\mu$m, i.e. $\tilde{\xi}=5.88$,
and a strength $R$ which is measured in units of a frequency $\omega$. As $R$ has the physical
units $\mbox{(energy)}^2 \mbox{(length)}^3$, the dimensionless disorder strength can 
be estimated
by $\tilde{R}=(\omega / \omega_{\rm g})^2 \tilde{\xi}^3$.
A weak disorder potential, which only slightly perturbes the condensate in the 
Florence experiment,
corresponds to the frequency $\omega = 2 \pi \cdot 50$ Hz, yielding $\tilde{R}=200$. 
In Figure~\ref{fig2} the
normalized shift of the critical temperature $\Delta T_c / (T_c^{(0)} \tilde{R})$ is 
plotted as a function
of the particle number $N$ for a Gaussian correlation, so we conclude that the 
disorder-induced shift
in the critical temperature is in this case of the order of 3 \%. With this it is about
the same size as
the interaction-induced shift \cite{Giorgini} and should thus be measurable \cite{Aspect1}.
Furthermore, we read off from Figure~\ref{fig2} that this shift increases with a decreasing 
number of
particles in the Bose gas.\\
In this letter we have determined the disorder-induced shift of the critical temperature for a 
Bose gas
in a harmonic trap. By treating the analytically solvable cases of a Gaussian and a Lorentzian
correlation function, we have estimated that this disorder-induced shift should be measurable 
in laser speckle experiments. Note that applying our general perturbative results 
(\ref{Omega1}) and (\ref{MUC}) to
a homogeneous Bose gas and to a vanishing correlation length $\xi= 0$
reproduces the seminal Lopatin-Vinokur shift (\ref{Hom}). However, in this case
the interaction-induced shift of the critical temperature
suggests that additional nonperturbative contributions to the disorder-induced
shift will certainly occur.
First indications for this conjecture are provided by
a recent momentum-shell renormalization group calculation \cite{Zobay}. \\
This work was supported by the SFB/TR 12 of the German Research Foundation (DFG).

\begin{thebibliography}{100}
%
\bibitem{Cornell}
M.H. Anderson, J.R. Ensher, M.R. Matthews, C.E. Wieman, and E.A. Cornell,
Science {\bf 269}, 198 (1995).
%
\bibitem{Ketterle1}
K.B. Davis, M.-O. Mewes, M.R. Andrews, N.J. van Druten, D.S. Durfee, D.M. Kurn, and W. Ketterle,
Phys. Rev. Lett. {\bf 75}, 3969 (1995).
%
\bibitem{Andersen}
J.O. Andersen,
Rev. Mod. Phys. {\bf 76}, 599 (2004).
%
\bibitem{Baym1}
G. Baym, J.P. Blaizot, M. Holzmann, F. Lalo{\"e}, and D. Vautherin, 
Phys. Rev. Lett. {\bf 83}, 1703 (1999).
%
\bibitem{Kleinertbec}
H. Kleinert,
Mod. Phys. Lett. B {\bf 17}, 1011 (2003).
%
\bibitem{Boris2}
B. Kastening,
Phys. Rev. A {\bf 69}, 043613 (2004).
%
\bibitem{Svistunov}
V.A. Kashurnikov, N.V. Prokof{\'{}}ev, and B.V. Svistunov, 
Phys. Rev. Lett. {\bf 87} 120402, (2001).
%
\bibitem{Arnold1}
P. Arnold and G. Moore,
Phys. Rev. Lett. {\bf 87}, 120401 (2001). 
%
\bibitem{Giorgini}
S. Giorgini, L.P. Pitaevskii, and S. Stringari,
Phys. Rev. A {\bf 54}, R4633 (1996).
%
\bibitem{Aspect1}
F.~Gerbier, J.H.~Thywissen, S.~Richard, M.~Hugbart, P.~Bouyer, and A.~Aspect,
Phys. Rev. Lett. {\bf 92}, 030405 (2004).
%
\bibitem{Zobay4}
O. Zobay, G. Metikas, and H. Kleinert,
Phys. Rev. A {\bf 71}, 043614 (2005)
%
\bibitem{Arnold3} 
P. Arnold, G. Moore, and B. Tom\'{a}\v{s}ik, 
Phys. Rev. A {\bf 65}, 013606 (2002).
%
\bibitem{Arnold4}
P. Arnold and B. Tom\'{a}\v{s}ik,
Phys. Rev. A {\bf 64}, 053609 (2001). 
%
\bibitem{Davis}
M.J. Davis and P.B. Blakie,
Phys. Rev. Lett. {\bf 96}, 060404 (2006). 
%
\bibitem{Kao}
Y.-M. Kao and T.F. Jiang,
Phys. Rev. A {\bf 73}, 043604 (2006).
%
\bibitem{Inguscio1}
J.E. Lye, L. Fallani, M. Modugno, D.S. Wiersma, C. Fort, and M. Inguscio,
Phys. Rev. Lett. {\bf 95}, 070401 (2005).
%
\bibitem{Aspect2}
D. Cl{\'e}ment, A.F. Var{\'o}n, M. Hugbart, J.A. Retter, P. Bouyer, L. Sanchez-Palencia, 
D.M. Gangardt, G.V. Shlyapnikov, and A. Aspect, 
Phys. Rev. Lett. {\bf 95}, 170409 (2005). 
%
\bibitem{Inguscio2}
C. Fort, L. Fallani, V. Guarrera, J.E. Lye, M. Modugno, D.S. Wiersma, and M. Inguscio, 
Phys. Rev. Lett. {\bf 95}, 170410 (2005).
%
\bibitem{Schulte}
T. Schulte, S. Drenkelforth, J. Kruse, W. Ertmer, J. Arlt, K. Sacha, J. Zakrzewski,
and M. Lewenstein,
Phys. Rev. Lett.  {\bf 95}, 170411 (2005).
%
\bibitem{Lewenstein}
B. Damski, J. Zakrzewski, L. Santos, P. Zoller, and M. Lewenstein,
Phys. Rev. Lett. {\bf 91}, 080403 (2003).
%
\bibitem{Castin}
U. Gavish and Y. Castin,  
Phys. Rev. Lett. {\bf 95}, 020401 (2005). 
%
\bibitem{Sengstock}
S. Ospelkaus, C. Ospelkaus, O. Wille, M. Succo, P. Ernst, K. Sengstock, and K. Bongs,
Phys. Rev. Lett. {\bf 96}, 180403 (2006).
%
\bibitem{Demler}
D.-W. Wang, M. D. Lukin, and E. Demler,
Phys. Rev. Lett. {\bf 92}, 076802 (2004).
%
\bibitem{Aspect3}
T. Schumm, J. Esteve, C. Figl, J.-B. Trebbia, C. Aussibal, H. Nguyen,
D. Mailly, I. Bouchoule, C.I. Westbrook, and A. Aspect,
Eur. Phys. J. D {\bf 32}, 171 (2005).
%
\bibitem{Schmiedmayer}
R. Folman, P. Kr{\"u}ger, J. Schmiedmayer, J. Denschlag, and C. Henkel,
Adv. At. Mol. Opt. Phys. {\bf 48}, 263 (2002).
%
\bibitem{Speckles}
J.C. Dainty (Ed.), {\it Laser Speckle and Related Phenomena} (Springer, Berlin, 1975).
%
\bibitem{Lopatin}
A.V. Lopatin and V.M. Vinokur,
Phys. Rev. Lett. {\bf 88}, 235503 (2002).
%
\bibitem{Reppy1}
B.C. Crooker, B. Hebral, E.N. Smith, Y. Takano, and J.D. Reppy,  
Phys. Rev. Lett. {\bf 51}, 666 (1983).
%
\bibitem{Reppy2}
M.H.W. Chan, K.I. Blum, S.Q. Murphy, G.K.S. Wong, and J.D. Reppy,
Phys. Rev. Lett. {\bf 61}, 1950 (1988).
%
\bibitem{Huang}
K. Huang and H.F. Meng,
Phys. Rev. Lett. {\bf 69}, 644 (1992). 
%
\bibitem{Graham}
R. Graham and A. Pelster,
eprint: {\tt cond-mat/0508306}.
%
\bibitem{Zobay}
O. Zobay,
Phys. Rev. A {\bf 73}, 023616 (2006).
%
\bibitem{Path}
H. Kleinert,
{\it Path Integrals in Quantum Mechanics, Statistics, Polymer Physics,
and Financial Markets},
Forth Extended Edition (World Scientific, Singapore, 2006).
%
\bibitem{Heitler}
W. Heitler, {\it The Quantum Theory of Radiation},
Third Edition (Oxford University Press, Oxford, 1970).
%
\bibitem{Zinn-Justin}
J. Zinn-Justin, {\it Quantum Field Theory and Critical Phenomena},
3rd Edition (Oxford University Press, Oxford, 1996).
%
\bibitem{Verena}
H.~Kleinert and V.~Schulte-Frohlinde, 
{\it Critical Properties of $\Phi^4$-Theories}
(World Scientific, Singapore, 2001).
%
\bibitem{Kling}
S. Kling and A. Pelster,
eprint: {\tt cond-mat/0604162}.
%
\bibitem{Dipol1} 
K. Glaum, A. Pelster, H. Kleinert, and T. Pfau,
eprint: {\tt cond-mat/0606569}.
%
\bibitem{Dipol2}
K. Glaum and A. Pelster,
eprint: {\tt cond-mat/0609374}.
%
\bibitem{Robinson}
J.E. Robinson,
Phys. Rev. {\bf 83}, 678 (1951).
%
\end{thebibliography}
\end{document}